\documentclass[letterpaper,12pt]{article}
\usepackage[margin=1.0in]{geometry}

\usepackage{amsmath,amssymb,graphicx,multirow,xspace,slashed}
\usepackage[usenames,dvipsnames]{xcolor}
\definecolor{DarkGreen}{rgb}{0,0.6,0}
\definecolor{DarkBlue}{rgb}{0,0,0.6}
\usepackage[colorlinks=true,linktocpage=true,linkcolor=DarkBlue,citecolor=DarkGreen,urlcolor=DarkGreen]{hyperref}
\usepackage[compress,numbers]{natbib}
\usepackage{subfigure, placeins}
\usepackage{booktabs}
\usepackage{blindtext, rotating}
\usepackage{afterpage}
\usepackage{enumitem}
\usepackage{ marvosym }
\usepackage{authblk} 
\usepackage{verbatim}
\usepackage{soul} %scratch out text
\usepackage[normalem]{ulem}
\usepackage{pifont}% http://ctan.org/pkg/pifont
\usepackage[usestackEOL]{stackengine}
\usepackage{subfiles}
\newcommand\sq{\framebox(10,10){}\kern\fboxrule}

\setstackgap{S}{\fboxrule}

\usepackage[utf8]{inputenc}

\usepackage{adjustbox}

\usepackage{floatrow}
\newfloatcommand{capbtabbox}{table}[][\FBwidth]

\definecolor{darkgreen}{rgb}{0,0.5,0}
\definecolor{goodyellow}{rgb}{0.9,0.7,0}

\usepackage[font=footnotesize,labelfont=bf]{caption}

\usepackage{lineno}

\allowdisplaybreaks

\renewcommand{\baselinestretch}{1.2}

\usepackage{bm}

\DeclareRobustCommand{\Sec}[1]{Sec.~\ref{#1}}

\DeclareRobustCommand{\Tab}[1]{Table~\ref{#1}}

\DeclareRobustCommand{\Fig}[1]{Fig.~\ref{#1}}
\DeclareRobustCommand{\Figs}[2]{Figs.~\ref{#1} and \ref{#2}}
\DeclareRobustCommand{\Eq}[1]{Eq.~\eqref{#1}}

\DeclareRobustCommand{\Ref}[1]{Ref.~\cite{#1}}
\DeclareRobustCommand{\Refs}[1]{Refs.~\cite{#1}}

\newcommand{\be}{\begin{equation}}
\newcommand{\ee}{\end{equation}}
\newcommand{\ba}{\begin{align}}
\newcommand{\ea}{\end{align}}
\usepackage{titlesec}

\newcommand{\ambone}{{
    \includegraphics[scale=0.04]{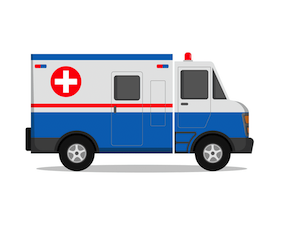}
}}
\newcommand{\ambtwo}{{
    \includegraphics[scale=0.04]{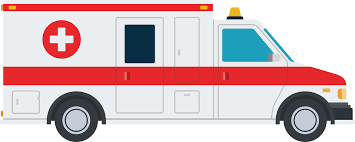}
}}

\newcommand{\mw}{$M_W$}

\title{
\vspace{-1.5cm}
\hspace{4.9in}
\small{MIT-CTP/5420}
\vspace{1.5cm}
\vspace{-0cm} \bf \Large
Oblique Lessons from the $W$ Mass Measurement at CDF~II
}

\author{Pouya Asadi$^{\ambone}$, Cari Cesarotti$^{\ambtwo}$, Katherine Fraser$^{\ambtwo}$, Samuel Homiller$^{\ambtwo}$, Aditya Parikh$^{\ambtwo}$

\vspace{0.5cm}

{\small $^{\ambone}$ Center for Theoretical Physics, \vspace{-0.5cm} Massachusetts Institute of Technology, \\
Cambridge, MA 02139, USA} \\ \vspace{-0.25cm}
{\small $^{\ambtwo}$ Department of Physics, Harvard University, Cambridge, MA, 02138, USA}}

\date{}

\begin{document}

\maketitle

\vspace{-0.5cm}

\begin{abstract}
The CDF collaboration recently reported a new precise measurement of the $W$ boson mass $M_W$ with a central value significantly larger than the SM prediction. 
We explore the effects of including this new measurement on a fit of the Standard Model (SM) to electroweak precision data. 
We characterize the tension of this new measurement with the SM and explore potential beyond the SM phenomena within the electroweak sector in terms of the oblique parameters $S$, $T$ and $U$. 
We show that the large $M_W$ value can be accommodated in the fit by a large, nonzero value of $U$, which is difficult to construct in explicit models.
Assuming $U = 0$, the electroweak fit strongly prefers large, positive values of $T$. Finally, we study how the preferred values of the oblique parameters may be generated in the context of models affecting the electroweak sector at tree- and loop-level. 
In particular, we demonstrate that the preferred values of $T$ and $S$ can be generated with a real SU(2)$_L$ triplet scalar, the humble \textit{swino}, which can be heavy enough to evade current collider constraints, or by (multiple) species of a singlet-doublet fermion pair. 
We highlight challenges in constructing other simple models, such as a dark photon, for explaining a large $M_W$ value, and several directions for further study.

\end{abstract}

\newpage
\tableofcontents

\section{Introduction}

The Standard Model of particle physics (SM) has been remarkably successful in explaining various experimental results. The discovery of the Higgs boson \cite{ATLAS:2012yve, CMS:2012qbp} at the Large Hadron Collider (LHC) was imperative to confirming the pattern of spontaneous symmetry breaking in the electroweak sector of the SM. However, as we continue to collect data and improve analysis techniques, we have seen a proliferation of precision measurements that deviate from SM predictions, such as the muon magnetic moment \cite{Muong-2:2021vma, Muong-2:2021ojo,Muong-2:2006rrc} and the $R_K/R_K^*$ anomalies \cite{LHCb:2017avl, LHCb:2019hip, LHCb:2021trn}. The most recent anomalous measurement reported is the mass of the $W$ boson \mw~\cite{CDF:2022hxs}. A discrepant measurement of \mw~could be an indication of supersymmetry (SUSY), composite Higgs, or other phenomena beyond the Standard Model (BSM) at potentially very high energy scales. It is therefore essential that we explore the phenomenological implications of this new \mw~measurement.

Measuring the $W$ mass with high precision requires a global fit of the SM, known as the electroweak fit. This method involves fitting over a set of well-measured SM observables, and minimizing the $\chi^2$ value over both the fitted (`free') observables as well as derived observables, see Refs.~\cite{Flacher:2008zq,Baak:2011ze,Baak:2014ora}. 
The electroweak fit leverages the small uncertainties of the fitted observables to produce precise predictions of the derived observables. 
Additionally, since this fit is an exceptional probe of precision measurements, it is also highly sensitive to BSM effects. 

We can parameterize the effects of new physics phenomena on the electroweak sector using oblique parameters $S$, $T$, and $U$ \cite{Peskin:1991sw, Peskin:1990zt} (see also Refs.~\cite{Kennedy:1988sn, Holdom:1990tc, Golden:1990ig}). 
These parameters capture the effects of higher-dimension operators \cite{Han:2004az, Han:2008es} that can arise in a variety of UV completions. In most models, $S$ and $T$ are the dominant corrections since they arise from dimension 6 operators, whereas $U$ is dimension 8 and therefore suppressed by a factor of ${v^2}/{\Lambda_\text{UV}^2}$.

The power of the electroweak fit is dependent on precision of experimental measurements of SM observables, and improves along with collider technology and luminosity. The leading measurements are made at the Large Electron-Positron Collider (LEP), Stanford Linear Collider (SLC), Tevatron, and  LHC. The discovery of the Higgs greatly improved the electroweak fit as it provided the final measured value to span the free parameters of the SM \cite{Zyla:2020zbs, Ciuchini:2013pca, deBlas:2016ojx}. 

The most recent update to the SM values used in the fit comes from the CDF collaboration at the Tevatron \cite{CDF:2022hxs}. Their analysis was completed with a four-fold increase of data, reduced uncertainty in PDFs and track reconstruction, and updated measurements compared to their previous result \cite{CDF:2013bqv}. They report 
\begin{equation}
 M_{W, \text{CDF~II}} = 80.4335 \pm 0.0094 \text{ GeV}, 
 \label{eq:CDFmw}
\end{equation}
which, without averaging with other experimental results, shows a 7$\sigma$ deviation from the SM prediction. This value is notably higher than the previous measurement averaged from the Tevatron and LEP experiments ($M_{W} = 80.385 \pm 0.015$ GeV) \cite{CDF:2013dpa}, as well as ATLAS ($M_W = 80.370 \pm 0.019$ MeV) \cite{ATLAS:2017rzl} and LHCb\footnote{This total error is calculated assuming all errors are uncorrelated and can be added in quadrature. The LHCb collaboration reports $M_W = 80.354 \pm 0.023_\text{stat} \pm 0.01_\text{exp} \pm 0.017_\text{th} \pm 0.009_\text{PDF}$.} ($M_W = 80.354 \pm 0.032$ GeV) \cite{LHCb:2021bjt}. 

In this paper we explore how new physics contributions, parameterized by the values of the oblique parameters, can adjust the electroweak fit such that \mw~is consistent with the updated CDF measurement. We first perform our fits scanning over values of $S$ and $T$ with $U$ fixed to zero (since $U$ is suppressed) and identify the range of these variables that can resolve the observed anomaly in \mw. We then study how the fit changes if we allow $U$ to float. Large values of $U$ can easily accommodate the observed increase in \mw; however, it is difficult to construct models with the primary new physics contributions affecting only $U$ while leaving $S$ and $T$ unchanged.

Next we consider several well-motivated simple extensions of SM that can produce nonzero $S$ and $T$ values. The models discussed in this paper include a generic dark photon with kinetic mixing, a two Higgs doublet model (2HDM), a neutral scalar SU(2)$_L$ triplet (that can be referred to correctly as a \textit{swino}), and various singlet-doublet fermion scenarios. For each model we check if there is available parameter space that corresponds to the fitted values of $T$ and $S$. We find that a dark photon or a scalar singlet/doublet  extension of SM can not explain the observed anomaly in \mw~measurements, while singlet-doublet fermion extension are strongly constrained by various experimental bounds. A $\mathcal{O}(\mathrm{TeV})$ swino, on the other hand, can explain the observed anomaly while evading current bounds and provides a well-motivated target for future high energy colliders.

The remainder of the paper is organized as follows. In \Sec{sec:EWPT}, we define the parameters and methodology of our electroweak fit. \Sec{sec:results} discusses the results and the implications of the oblique parameters on fitting the measured observables. In \Sec{sec:bsm} we map the values of the fitted oblique parameters to the parameters of various models, and comment on the viability of this space. We conclude in \Sec{sec:conc}.

\section{Electroweak Fit}
\label{sec:EWPT}

To assess the impact of the new measurements of $M_W$, and the implications for potential new physics, we perform an electroweak fit to a representative set of observables, following the strategy of the GFitter group~\cite{Flacher:2008zq, Baak:2011ze, Baak:2014ora, Haller:2018nnx}\footnote{
With respect to the GFitter results in \Ref{Haller:2018nnx}, we consider an updated value of the Higgs mass and the revised values of $\Gamma_Z$ and $\sigma^0_{\textrm{had}}$ from \Ref{Janot:2019oyi}.} with a modified version of the code used in Refs.~\cite{Fan:2014vta, Fan:2014axa}.
A set of five core observables are free to vary in the fit: the $Z$ boson mass $M_Z$, the top mass $M_t$, the Higgs mass $M_h$, the $Z$-pole value of the strong coupling constant $\alpha_s(M_Z)$, and the hadronic contribution to the running of $\alpha$, denoted $\Delta \alpha_{\textrm{had}}^{(5)}(M_Z^2)$.

These five predicted values and other observables derived from them are compared to their measured values (see \Tab{tab:obs_table}). 
In addition to measurements of these five parameters, the observables considered include the $W$ mass and a host of other electroweak precision measurements performed at SLC, LEP, the Tevatron, and the LHC, which are listed with their measured values below the horizontal line in \Tab{tab:obs_table}. 
These other observables can be determined in the SM as functions of the five core observables, the Fermi constant $G_F$, and the fine structure constant $\alpha(q^2 = 0)$. In the electroweak fit, $G_F = 1.1663787\,\times\,10^{-5}\,\textrm{GeV}^{-2}$ and $\alpha = 1/137.03599084$ are treated as fixed values since they are determined with much higher precision than the rest of the observables \cite{Zyla:2020zbs}. 

\begin{table}[]
    \centering
    \begin{tabular}{l |c}
         Observable & Measured Value 
         \\ \hline
         $M_Z$ [GeV]								& $91.1876 \pm 0.0021$ \\
         $M_h$ [GeV]								& $125.25 \pm 0.17$ \\
         $M_t$ [GeV]									& $172.89 \pm 0.59$ \\
         $\alpha_s(M_Z^2)$ 						& $0.1181 \pm 0.0011$ \\
         $\Delta \alpha_{\textrm{had}}^{(5)}(M_Z^2)$ & $0.02766 \pm 0.00007$	
         \\[0.25em]
 		\hline
         $\Gamma_Z$ [GeV]						& $2.4955 \pm 0.0023$ \\
         %$\Gamma_{\textrm{inv.}}$	[GeV] 	& $0.499 \pm 0.0025$ \\
         $\Gamma_W$ [GeV] 					& $2.085 \pm 0.042$ \\
         $\sigma^0_{\textrm{had}}$~[nb] 	& $41.481 \pm 0.033$ \\
         $R^0_{\ell}$ 									& $20.767 \pm 0.025$ \\
         $A_{\textrm{FB}}^{0,\ell}$				& $0.0171 \pm 0.0010$ \\
         $A_{\ell}$ 									& $0.1499 \pm 0.0018$ \\
         $\sin^2\theta^{\ell}_{\textrm{eff}}(Q_{\textrm{FB}})$ &	$0.2324 \pm 0.0012$ \\
         $\sin^2\theta^{\ell}_{\textrm{eff}}(\textrm{Tevt.})$ 	& $0.23148 \pm 0.00033$ \\
         $A_b$ 											& $0.923 \pm 0.020$ \\
         $A_c$ 											& $0.670 \pm 0.027$ \\
         $A^{0,b}_{\textrm{FB}}$ 				& $0.0992 \pm 0.0016$ \\
         $A^{0,c}_{\textrm{FB}}$ 				& $0.0707 \pm 0.0035$ \\
         $R^{0,b}$ 									& $0.21629 \pm 0.00066$ \\
         $R^{0,c}$ 									& $0.1721 \pm 0.0030$ \\
         \hline
    \end{tabular}
    \caption{
    Summary of the observables included in the fit, and their experimental values. 
    The five observables above the horizontal line are allowed to float in the fit, 
	while the SM values of the remaining observables are determined from these five values, as discussed in the main text.
    The values of $M_Z$, $M_t$, $M_h$, $\alpha_s(M_Z^2)$, $\Delta \alpha_{\textrm{had}}^{(5)}(M_Z^2)$, and $\Gamma_W$ are taken from the most recent PDG average~\cite{Zyla:2020zbs}.
    For $\Gamma_Z$ and $\sigma^0_{\textrm{had}}$ we use the updated values computed in Ref.~\cite{Janot:2019oyi}.
    The remaining $Z$-pole observables are taken from the LEP and SLC measurements~\cite{ALEPH:2005ab}. 
    For $A_{\ell}$ we use the average of the LEP and SLC values, following \Ref{Haller:2018nnx}.}
    \label{tab:obs_table}
\end{table}

For the $W$ mass, we will consider several different values to assess the impact of the recent CDF measurement on the overall state of the global EW fit. These are, 
%boe%
\begin{equation}
\begin{aligned}
    \qquad M_W & = 80.4335 \pm 0.0094 ~\textrm{GeV} \qquad \text{(CDF~II)}, \\
%    \qquad M_W & = 80.4242 \pm 0.0087 ~\textrm{GeV} \qquad \text{(Tevatron + LEP)} \\
    \qquad M_W & = 80.4112 \pm 0.0076 ~ \textrm{GeV} \qquad \text{(LHC + LEP + Tevatron)}, \\
    \qquad M_W & = 80.379 \pm 0.012 ~ \textrm{GeV}\phantom{00} \qquad \text{(PDG 2020)}, 
    \label{eq:Wscenarios}
\end{aligned}
\end{equation}
%eoe%
where the uncertainties quoted above include the statistical, systematic and modeling uncertainties used in each experiment. 
The second scenario is our estimate for the global average of different \mw~measurements, assuming  zero correlations between experimental result to first approximation.\footnote{While there are sources of uncertainty such as parton distribution functions that might introduce some correlation between these results, when we repeated the world average \mw~scenario (Tevatron + LEP + LHC) with a few different values for the correlations, we arrived at similar qualitative results. A comprehensive global averaging of these experimental results considering all correlations is left for future work.}
In addition, to assess the particular impact of the new, high precision measurement from CDF~II, we will also perform the fit with $M_W$ taken to be the CDF~II value with the systematic uncertainty artificially inflated by a factor of 2, $M_W = 80.4335 \pm 0.0157$, to better understand the compatibility of the CDF measurement with the SM prediction. This scenario is referred to as the CDF~II (2x Syst.) throughout the paper.

The SM values of the other observables are determined from the free parameters using the full two-loop electroweak results available in the literature. 
The running of $\alpha$ is computed using the floating value of $\Delta \alpha_{\textrm{had}}^{(5)}$ as well as the leptonic piece, $\Delta \alpha_{\textrm{lep}} = 0.031497686$~\cite{Steinhauser:1998rq}, which is kept fixed in the fit.
The $W$ mass is determined using the parameterization in \Ref{Awramik:2003rn}, which also includes corrections up to $\mathcal{O}(\alpha \alpha_s^3)$ for the radiative correction (referred to as $\Delta r$ in the literature).
The expression for the width of the $W$ is taken from the parameterization in \Ref{Cho:2011rk}.
For the $Z$ width $\Gamma_Z$, hadronic peak cross section $\sigma^0_{\textrm{had}}$, and width ratios $R^0_{\ell}$, $R^0_b$, $R^0_c$, we use the parameterizations in \Ref{Dubovyk:2018rlg}.
For the effective weak mixing angle, $\sin^2\theta_{\textrm{eff}}^{\ell}$, we use the results in \Ref{Awramik:2006uz}.
The value of $\sin^2\theta_{\textrm{eff}}^{\ell}$ is used as a proxy for the weak mixing angle to determine the left- and right-handed couplings of the $Z$, allowing us to compute the asymmetries:
%boe%
\begin{equation}
\mathcal{A}_f = \frac{ g_{Lf}^2 - g_{Rf}^2}{g_{Lf}^2 + g_{Rf}^2}
\end{equation}
%eoe%
for $f = \ell, c, b$. The value of $\sin^2\theta_{\textrm{eff}}^{\ell}$ is also used to compute the forward-backward asymmetry $A_{\textrm{FB}}^{0,\ell}$.
Finally, for the other forward-backward asymmetries, we compute the effective weak mixing angles $\sin^2\theta_{\textrm{eff}}^{b}$ and $\sin^2\theta_{\textrm{eff}}^{c}$ using the parameterizations in Refs.~\cite{Awramik:2006uz,Dubovyk:2016aqv}, respectively. These are then translated to $A_{\textrm{FB}}^{0, b, c}$ using the standard relations summarized e.g., in \Ref{Dubovyk:2016aqv}. See also Ref.~\cite{Erler:2019hds} for a recent review of the status of relevant theoretical calculations.

We parameterize potential effects of BSM physics in the electroweak fit in terms of the oblique parameters, $S$, $T$ and $U$~\cite{Peskin:1991sw, Peskin:1990zt}: 
\begin{equation}
\begin{split}
%boe%
S & \equiv \frac{4c_{W}^2 s_{W}^2}{\alpha} \left[ \Pi'_{ZZ}(0) -\frac{c_{W}^2 -s_{W}^2}{c_{W}s_{W}}  \Pi_{Z\gamma}'(0) -\Pi_{\gamma\gamma}'(0) \right], \\[0.25em]
T & \equiv \frac{1}{\alpha} \left[ \frac{\Pi_{WW}(0)}{m_W^2} -\frac{\Pi_{ZZ}(0)}{m_Z^2} \right],\\[0.25em]
U & \equiv \frac{4s_{W}^2}{\alpha} \left[ \Pi'_{WW}(0) -\frac{c_{W}}{s_{W}}  \Pi_{Z\gamma}'(0) -\Pi_{\gamma\gamma}'(0) \right] -S,
\end{split}
\label{eqn:STU}
\end{equation}
%eoe%
where $\Pi_{XX}$ denotes the vacuum polarization for $X = W, Z, \gamma$, and $c_W$, $s_W$ are $\cos\theta_W$, $\sin\theta_W$ with $\theta_W$ denoting the Weinberg mixing angle. 
(Note that $S$, $T$ and $U$ do not completely characterize potential BSM effects in the electroweak precision data---a larger set of oblique parameters was developed in Refs.~\cite{Barbieri:2004qk, Cacciapaglia:2006pk}. We will not consider their effects here, as they are typically smaller in perturbative theories~\cite{Fan:2014vta, Cho:1994yu}.) 
The new physics contributions to the electroweak observables can be expressed as linear functions of $S$, $T$ and $U$~\cite{Peskin:1991sw, Peskin:1990zt, Maksymyk:1993zm,Burgess:1993mg,Burgess:1993vc}, which are are summarized in Appendix~A of \Ref{Ciuchini:2013pca}.

For a class of universal effective theories, both $S$ and $T$ are related to the Wilson coefficients \cite{Han:2004az, Han:2008es, Grzadkowski:2010es}  of dimension-6 operators\footnote{See Ref.~\cite{Wells:2015uba} for a detailed discussion of the relationship between the oblique parameters and effective theories.}: 
\begin{equation}
    \mathcal{L}_\text{oblique} = S \left( \frac{\alpha}{s_W c_W v^2 }\right)H^\dagger W_a^{\mu\nu}\sigma^a H B_{\mu\nu} - T \left( \frac{2\alpha}{v^2}\right)|H^\dagger D_\mu H|^2.
\label{eq:effOper}
\end{equation}
The $U$ parameter is often fixed to zero in electroweak fits, as it corresponds to a dimension-8 operator from an effective field theory point of view, and its effects are therefore subleading compared to $S$ and $T$. 
We will frequently set $U = 0$ in our fits, but consider its effect in more details in \Sec{subsec:u_param}. 
We will discuss new physics interpretations of $S$ and $T$ following the results of the fit with $U = 0$ in \Sec{sec:bsm}.

With all of these inputs, we perform the electroweak fit by minimizing a $\chi^2$ function,
%boe%
\begin{equation}
\chi^2 = \sum_j \bigg( \frac{ M_j - O_j }{\sigma_j} \bigg)^2 \, ,
\end{equation}
%eoe%
where the sum runs over all the observables in Table~\ref{tab:obs_table}, in addition to the $W$ mass. 
Here, $M_j$ is the experimentally measured value of the observable, $O_j$ is the predicted value in terms of the five free parameters and $S, T, U$, and $\sigma_j$ is the measured uncertainty on the observable. 
We repeat this calculation for all the four scenarios for \mw~measurements defined around Eq.~\eqref{eq:Wscenarios}.

%============================================================
\section{Results of the Fit}
\label{sec:results}

\subsection{Fitting $S$ and $T$}

We first consider the fit results where $U$ is fixed to zero.
The results of our electroweak fit with different values of $M_W$ are summarized in \Tab{tab:summary}.
The first row indicates the $\chi^2$ per degree of freedom (d.o.f.) for the SM for the fit with each value of $M_W$. We observe that, prior to the CDF measurement, the Standard Model provides a good fit to the data using the PDG 2020 value of $M_W$, with $\chi^2 / (n_{\textrm{d.o.f.}} = 15)=1$ ($p = 0.45$).
Taking instead the recent CDF~II measurement of $M_W$, however, the $p$-value for the SM drops to $2.52 \times 10^{-7}$, exemplifying the tension discussed in \Ref{CDF:2022hxs}.
This is somewhat ameliorated when considering the smaller world average value of $M_W$ ($p = 1.06 \times 10^{-4}$), but notable tension remains.

In the middle rows of \Tab{tab:summary} we summarize the results of the fit when we allow $S$ and $T$ to float in addition to the five free observables. We report the best fit values of $S$ and $T$, and then the $\chi^2$ per degree of freedom. We find a good fit to the data with the PDG average value of $M_W$, prior to the CDF measurement ($p = 0.53$), where the fit prefers small values of $S$ and $T$ at $0.05$ and $0.08$, respectively. This is consistent with the electroweak fit presented in \Ref{Zyla:2020zbs}.
For all of the fits accounting for the new measurement of $M_W$ from CDF~II, the fit instead prefers much larger values of $S$ and $T$. Despite this, we still find a good fit to the data, with $p$ values ranging from $0.26$ when using the CDF measurement alone to $0.41$ using the combination of measurements at LHC, Tevatron, and LEP.

%%%%%%%%%%
\begin{table}[t]
\renewcommand{\arraystretch}{1.2}
\resizebox{\columnwidth}{!}{
\begin{tabular}{lc|cccc}
\hline
 &  & \multirow{2}{*}{CDF-II} & CDF-II & \multirow{2}{*}{World Average} & \multirow{2}{*}{PDG 2020} \\
 &  &  & \multicolumn{1}{c}{($2\times$ syst.)} &  &  \\
 \hline
%$\chi^2$ (SM)   & 59.9788 & 34.4369 & 44.0319 & 14.9685     \\
SM & $\chi^2 /(n_{\textrm{d.o.f.}} = 15)$ 
                & 4.00 &    2.30 &  2.94 & 1.00 \\ \hline
%\multirow{2}{*}{Best Fit} & $(S, T)$ 
%               & (0.17, 0.27)  & (0.15, 0.24)  & (0.12, 0.20)  & (0.05, 0.08) \\
% & ($\chi^2$)  & 15.7002       & 15.2435       & 13.5265       & 11.9114      \\
\multirow{2}{*}{Best Fit ($U=0$)} & $(S, T)$ 
                & (0.17, 0.27) & (0.15, 0.24) & (0.12, 0.20)    & (0.05, 0.08) \\
  & $\chi^2 / (n_{\textrm{d.o.f}} = 13)$ 
                & 1.21         & 1.17         & 1.04            & 0.92 
\\\hline
%\multirow{2}{*}{Best Fit} & $(S, T, U)$ & (0.03, 0.05, 0.19) & (0.03, 0.05, 0.19) & (0.03, 0.05, 0.12) & (0.03, 0.05, 0.03) \\  & $\chi^2 / (n_{\textrm{d.o.f}} = 12)$ & 11.8746 & 11.8727 & 11.8518 & 11.829
\multirow{2}{*}{Best Fit ($U$ floating)} & $(S, T, U)$ & (0.03, 0.05, 0.19) & (0.03, 0.05, 0.19) & (0.03, 0.05, 0.12) & (0.03, 0.05, 0.03) \\  & $\chi^2 / (n_{\textrm{d.o.f}} = 12)$ & 0.99 & 0.99 & 0.99 & 0.99
\\\hline
\end{tabular}
}
\caption{Fit results including the oblique parameters and $\chi^2$ per degree of freedom. Different columns correspond to different input \mw~measurement scenarios around Eq.~\eqref{eq:Wscenarios}.The first row shows the $\chi^2$ per degree of freedom for SM in each \mw~scenario. Results of the fit including (excluding) $U$ in the list of floating parameters are included in the middle (bottom) row. }
\label{tab:summary}
\end{table}
%%%%%%%%%%

\begin{figure}
    \centering
    \includegraphics[width=0.5\textwidth]{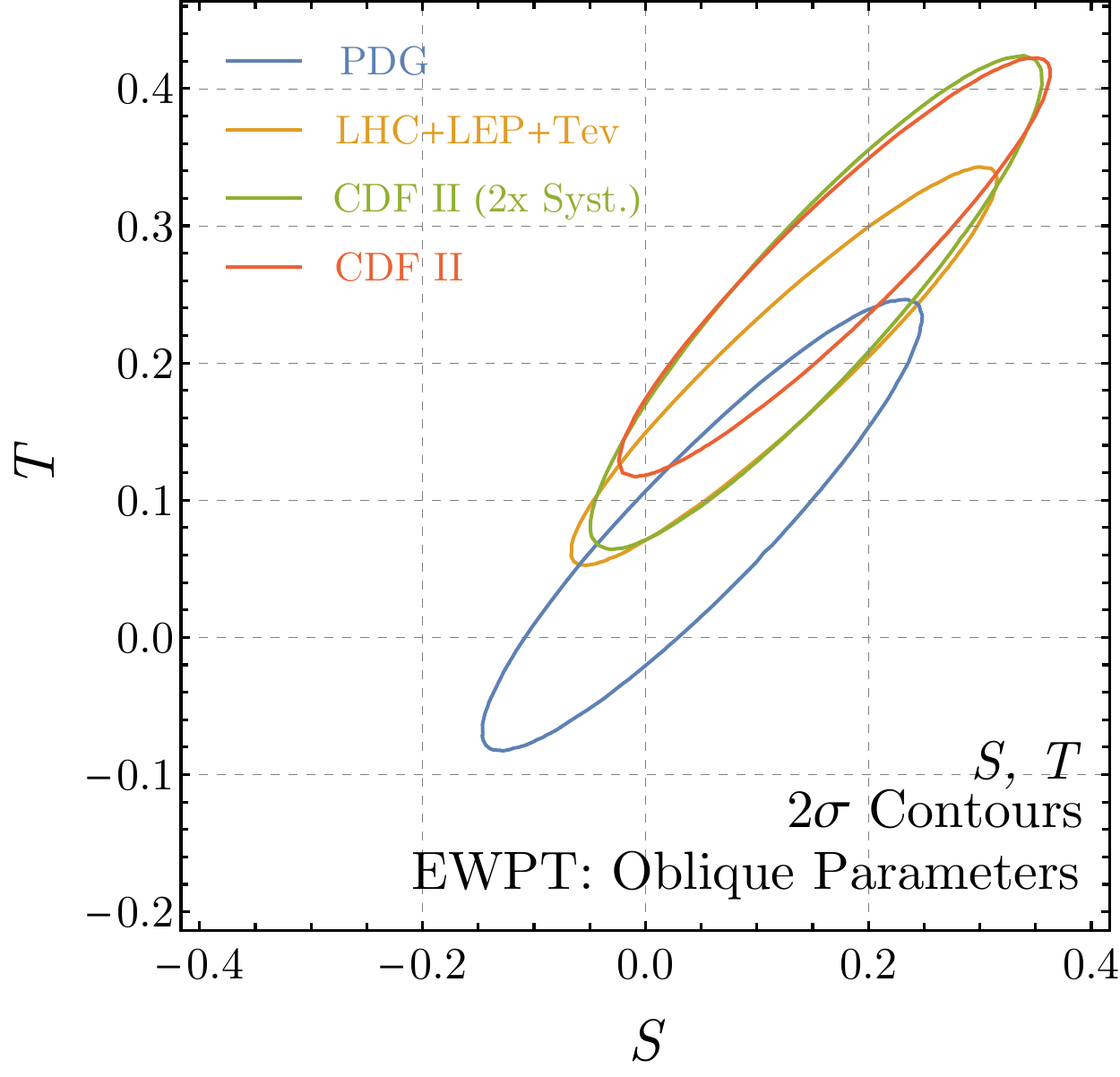}
    \caption{
    The $2\sigma$ preferred regions in the $S$ and $T$ plane from the electroweak fit, marginalizing over the five input parameters and for various experimental values of $M_W$ (see the discussion around Eq.~\eqref{eq:Wscenarios}). We do not include $U$ in these fits. The blue curve is in good agreement with results of GFitter group~\cite{Flacher:2008zq, Baak:2011ze, Baak:2014ora, Haller:2018nnx}. Including the recent CDF~II measurement of \mw~\cite{CDF:2022hxs} moves the best-fit region to larger positive values of $S$ and $T$. The SM (with $(S,~T)=(0,~0)$) is strongly disfavored when the new CDF~II \mw~measurement is included in the fit. 
    }
    \label{fig:STplane}
\end{figure}

The results of the fit for the oblique parameters $S$ and $T$ are illustrated in Fig.~\ref{fig:STplane}. Here we show ellipses indicating the $2\,\sigma$ contours around the best-fit values of $S$ and $T$. These are computed by computing the $\chi^2$ at each point in the $S-T$ plane, marginalizing over the free observables, and requiring $\Delta \chi^2 \equiv \chi^2(S,T) - \chi^2_{\textrm{min}} < 6.18$, where $\chi^2_{\textrm{min}}$ is the minimum value of the $\chi^2$ as a function of all the free parameters as well as $S$ and $T$.

%%%%%%%%%%
%%%%%%%%%%

The $2\sigma$ contours of the fit with the PDG average value of $M_W$ (excluding the recent CDF~II measurement) are shown in blue and agree with the results of \Ref{Haller:2018nnx}.
This fit slightly prefers $T > 0$, though the correlation between $S$ and $T$ leaves some parameter space with $S, T < 0$ as well.
Once the new measurement of $M_W$ from CDF~II is included, however, the preferred region in the $S-T$ plane shifts dramatically. The correlation between $S$ and $T$ remains, but values of $T < 0$ are no longer allowed, even when the systematic error on the CDF measurement is artificially inflated. 
In all, we find a strong preference for BSM contributions in the electroweak fit, particularly for positive, nonzero values of $T$.

For each fit, we also find the best fit value of each individual observable  both for the SM (with $S$ and $T$ fixed to zero) and for the best-fit value of $S$ and $T$. The results are shown in  Table~\ref{tab:observable_fit}. 
Each entry indicates the best-fit value of the observable, along with the pull (calculated as the fit value minus the measured value, divided by the experimental uncertainty) shown in parentheses.
For all three values of $M_W$ including the new CDF measurement, we see a significant pull (ranging from $-4.5$ to $-6.3$) on the fit value of $M_W$ in the Standard Model.
This is entirely ameliorated at the best fit values of $S$ and $T$, at the cost of a small tension in the values of $\Gamma_Z$, which are fit to be larger than the experimental value when $S$ and $T$ are allowed to float.
All of the other observables have quite similar values at their best-fit point and at the SM, regardless of the experimental value of $M_W$ used in the fit.
Note also that the previously existing tension in the forward-backward asymmetry, $A^{0,b}_{\textrm{FB}}$ measured at LEP is unaffected by the floated values of $S$ and $T$ and is roughly the same for any value of $M_W$.

%%%%%%%%%%
\begin{table}[hp!]
\begin{adjustbox}{angle=-90}
\resizebox{1.25\columnwidth}{!}{
\begin{tabular}{l|cc|cc|cc|cc}
\hline
\multicolumn{1}{c|}{} & \multicolumn{8}{c}{Fit Value (Pull) at SM / Best Fit (S,T)} \\
\multicolumn{1}{c|}{} & \multicolumn{2}{c}{CDF-II} & \multicolumn{2}{c}{CDF-II ($2\times$ syst.)} & \multicolumn{2}{c}{World Average} & 
\multicolumn{2}{c}{PDG 2020} \\
\hline
$(S,T)$ & (0,0) & (0.17, 0.27) & (0,0) & (0.15, 0.24) & (0,0) & (0.12, 0.20) & (0,0) &  (0.05, 0.08) \\
 \hline
$M_Z$ [GeV] & 
91.1910 (+1.6) & 91.1875 (0.0) & 91.1893 (+0.8) & 91.1876 (0.0) & 91.1908 (+1.5) & 91.1876 (0.0) & 91.1884 (+0.4) & 91.1876 (0.0) \\
$M_h$ [GeV] &
125.24 (-0.1) & 125.26 (+0.1) & 125.25 (0.0) & 125.25 (0.0) & 125.24 (-0.1) & 125.26 (+0.1) & 125.25 (0.0) & 125.25 (0.0) \\
$M_t$ [GeV] & 
174.14 (+2.1) & 172.93 (+0.1) & 173.49 (+1.0) & 172.90 (0.0) & 174.07 (+2.0) & 172.91 (0.0) & 173.16 (+0.46) & 172.93 (+0.1) \\
$\alpha_s(M_Z^2)$ & 
 0.1179 (-0.2) & 0.1179 (-0.2) & 0.1182 (+0.1) & 0.1179 (-0.2) & 0.1180 (-0.1) & 0.1180 (-0.1) & 0.1183 (+0.2) & 0.1183 (+0.2) \\
$\Delta\alpha_{\textrm{had}}^{(5)}(M_Z^2)$ & 
0.02761 (-0.7) & 0.02766 (0.0) & 0.02764 (-0.3) & 0.02766 (0.0) & 0.02761 (-0.7) & 0.02766 (0.0) & 0.02764 (-0.3) & 0.02766 (0.0) \\[0.25em] 
\hline
$M_W$ [GeV] & 
80.3690 (-6.3) & 80.4272 (-0.6) & 80.3622 (-4.5) & 80.4204 (-0.8) & 80.3682 (-5.0) & 80.4082 (-0.3) & 80.3587 (-1.6) & 80.3776 (-0.1) \\
$\Gamma_Z$ [GeV] & 
2.4950 (-0.2) & 2.4995 (+1.7) & 2.4948 (-0.3) & 2.4990 (+1.5) & 2.4950 (-0.2) & 2.4982 (+1.2) & 2.4946 (-0.4) & 2.4961 (+0.3) \\
$\Gamma_W$ [GeV] & 
2.091 (+0.1) & 2.096 (+0.3) & 2.091 (+0.1) & 2.095 (+0.2) & 2.091 (+0.1) & 2.094 (+0.2) & 2.090 (+0.1) & 2.092 (+0.2) \\
$\sigma^0_{\textrm{had.}}$ [nb] & 
41.489 (+0.2) & 41.492 (+0.3) & 41.489 (+0.2) & 41.492 (+0.3) & 41.489 (+0.2) & 41.491 (+0.3) & 41.489 (+0.2) & 41.490 (+0.3) \\
$R^0_{\ell}$ & 
20.748 (-0.8) & 20.749 (-0.7) & 20.750 (-0.7) & 20.749 (-0.7) & 20.748 (-0.8) & 20.749 (-0.7) & 20.750 (-0.7) & 20.750 (-0.7) \\
$A^{0,\ell}_{\textrm{FB}}$ & 
0.0163 (-0.8) & 0.0163 (-0.8) & 0.0162 (-0.9) & 0.0163 (-0.8) & 0.0163 (-0.8) & 0.0163 (-0.8) & 0.0162 (-0.9) & 0.0162 (-0.9) \\
$A_{\ell}$ & 
0.1474 (-1.4)  & 0.1474 (-1.4) & 0.1471 (-1.6) & 0.1473 (-1.4) & 0.1474 (-1.4) & 0.1472 (-1.5) & 0.1469 (-1.7) & 0.1470 (-1.6) \\
$\sin^2\theta^{\ell}_{\textrm{eff}}(Q_{\textrm{FB}})$ &
 0.2315 (-0.8) & 0.2315 (-0.8) & 0.2315 (-0.8) & 0.2315 (-0.8) & 0.2315 (-0.8) & 0.2315 (-0.8) & 0.2315 (-0.8) & 0.2315 (-0.8) \\
$\sin^2\theta^{\ell}_{\textrm{eff}}(\textrm{Tevt.})$ & 
 0.23147 (0.0) & 0.23148 (0.0) & 0.23152 (+0.1) & 0.23148 (0.0) & 0.23148 (0.0) & 0.23150 (+0.1) & 0.23154 (+0.2) & 0.23153 (+0.2) \\
$A_b$ & 
0.936 (+0.7)  & 0.936 (+0.7) & 0.936 (+0.7) & 0.936 (+0.7) & 0.936 (+0.7) & 0.936 (+0.7) & 0.936 (+0.7) & 0.936 (+0.7) \\
$A_c$ & 
0.668 (-0.1) & 0.668 (-0.1) & 0.668 (-0.1) & 0.668 (-0.1) & 0.668 (-0.1) & 0.668 (-0.1) & 0.667 (-0.1) & 0.667 (-0.1) \\
$A^{0,b}_{\textrm{FB}}$ & 
0.1033 (+2.6) & 0.1033 (+2.6) & 0.1031 (+2.4) & 0.1033 (+2.6) & 0.1033 (+2.6) & 0.1032 (+2.5) & 0.1030 (+2.4) & 0.1030 (+2.4) \\
$A^{0,c}_{\textrm{FB}}$ & 
0.0739 (+0.9)  & 0.0739 (+0.9) & 0.0737 (+0.9) & 0.0738 (+0.9) & 0.0738 (+0.9) & 0.0738 (+0.9) & 0.0736 (+0.8) & 0.0736 (+0.8) \\
$R^{0,b}$ & 
0.21582 (-0.7) & 0.21586 (-0.7) & 0.21584 (-0.7) & 0.21586 (-0.7) & 0.21583 (-0.7) & 0.21586 (-0.7) & 0.21585 (-0.7) & 0.21586 (-0.7) \\
$R^{0,c}$ & 
0.1722 (0.0) & 0.1722 (0.0) & 0.1722 (0.0) & 0.1722 (0.0) & 0.1722 (0.0) & 0.1722 (0.0) & 0.1722 (0.0) & 0.1722 (0.0) \\
\hline
\end{tabular}
}
\caption{The best fit values of the observables, and their pulls (calculated as the fit value minus the measured value, divided by the experimental uncertainty).}
\label{tab:observable_fit}
\end{adjustbox}
\end{table}

%%%%%%%%%%

%============================================================
\subsection{The $U$ Parameter}
\label{subsec:u_param}

In the fits described above, we have fixed $U = 0$.
As discussed in \Sec{sec:EWPT}, this is motivated by the fact that the $U$-parameter is dimension 8, and is typically suppressed relative to $S$ and $T$ in concrete models.

Nevertheless, in light of the large value of $M_W$ measured at CDF~II, it is worth examining the effects of the $U$-parameter on the electroweak fits in more detail. 
This is because, of all the electroweak precision observables we consider, the $U$ parameter affects only two: the $W$ mass and width~\cite{Maksymyk:1993zm,Burgess:1993mg,Ciuchini:2013pca}:\footnote{We thank Ayres Freitas for emphasizing this point to us.}
%boe%
\begin{equation}
\begin{aligned}
M_W & = M_{W,\textrm{SM}}
\bigg( 1 - \frac{\alpha(M_Z^2)}{4(c_W^2 - s_W^2)} 
\big( S - 2 c_W^2 T\big) + \frac{\alpha(M_Z^2)}{8 s_W^2} U \bigg), 
\\[0.25em]
\Gamma_W & = \Gamma_{W,\textrm{SM}} \bigg( 1 - \frac{3 \alpha(M_Z^2)}{4(c_W^2 - s_W^2)} 
\big( S - 2 c_W^2 T\big) + \frac{3 \alpha(M_Z^2)}{8 s_W^2} U \bigg) .
\end{aligned}
\end{equation}
%eoe%
The $W$ decay width is not measured to nearly as high precision as $M_W$, 
so the observed discrepancy in the $W$ mass at CDF~II \cite{CDF:2022hxs} can be accommodated in the SM electroweak fit by setting $U \approx 0.11$, without affecting any of the other observables.

\begin{figure}
    \centering
    \includegraphics[width=.3\textwidth]{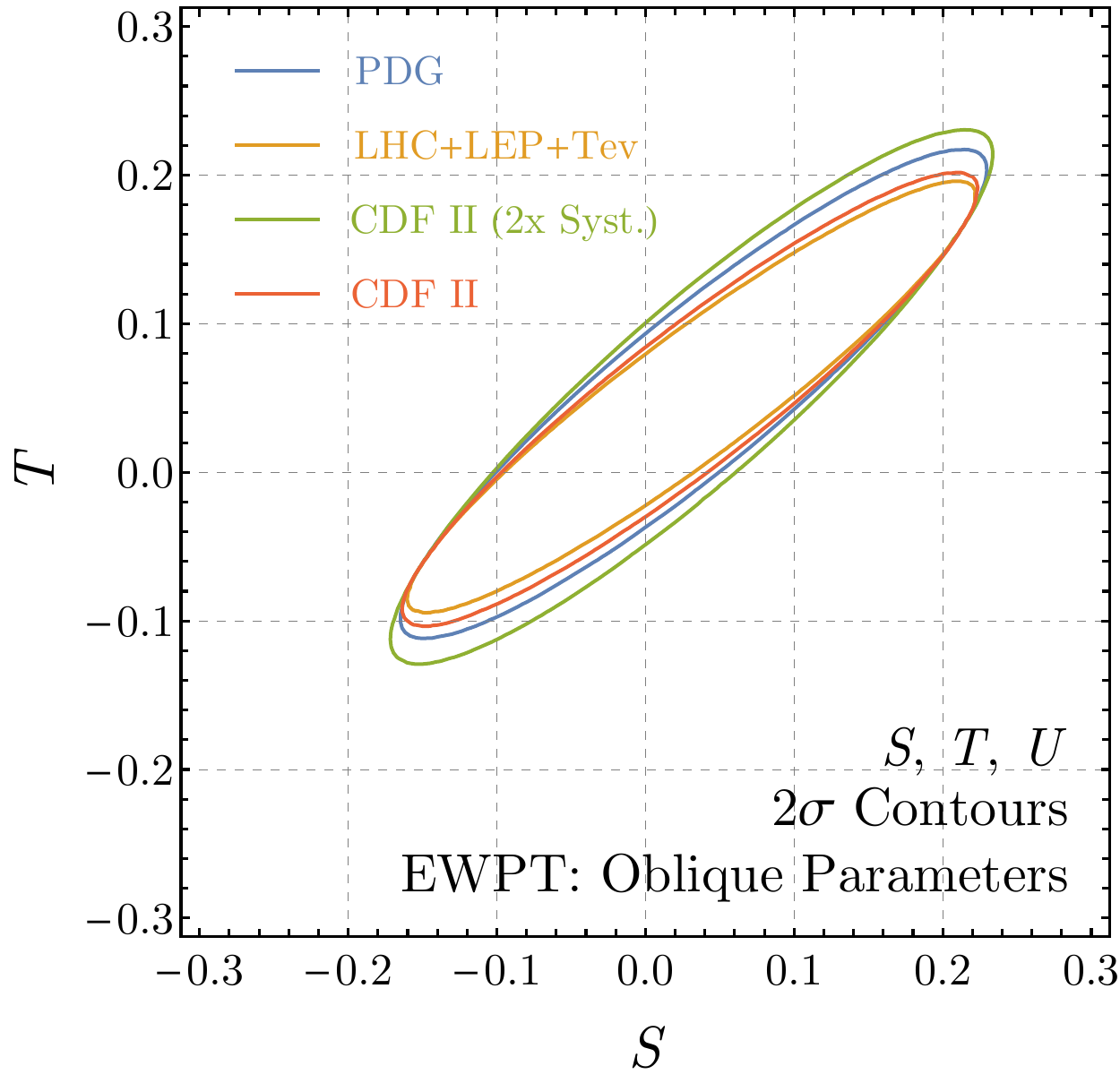}\quad 
    \includegraphics[width=.3\textwidth]{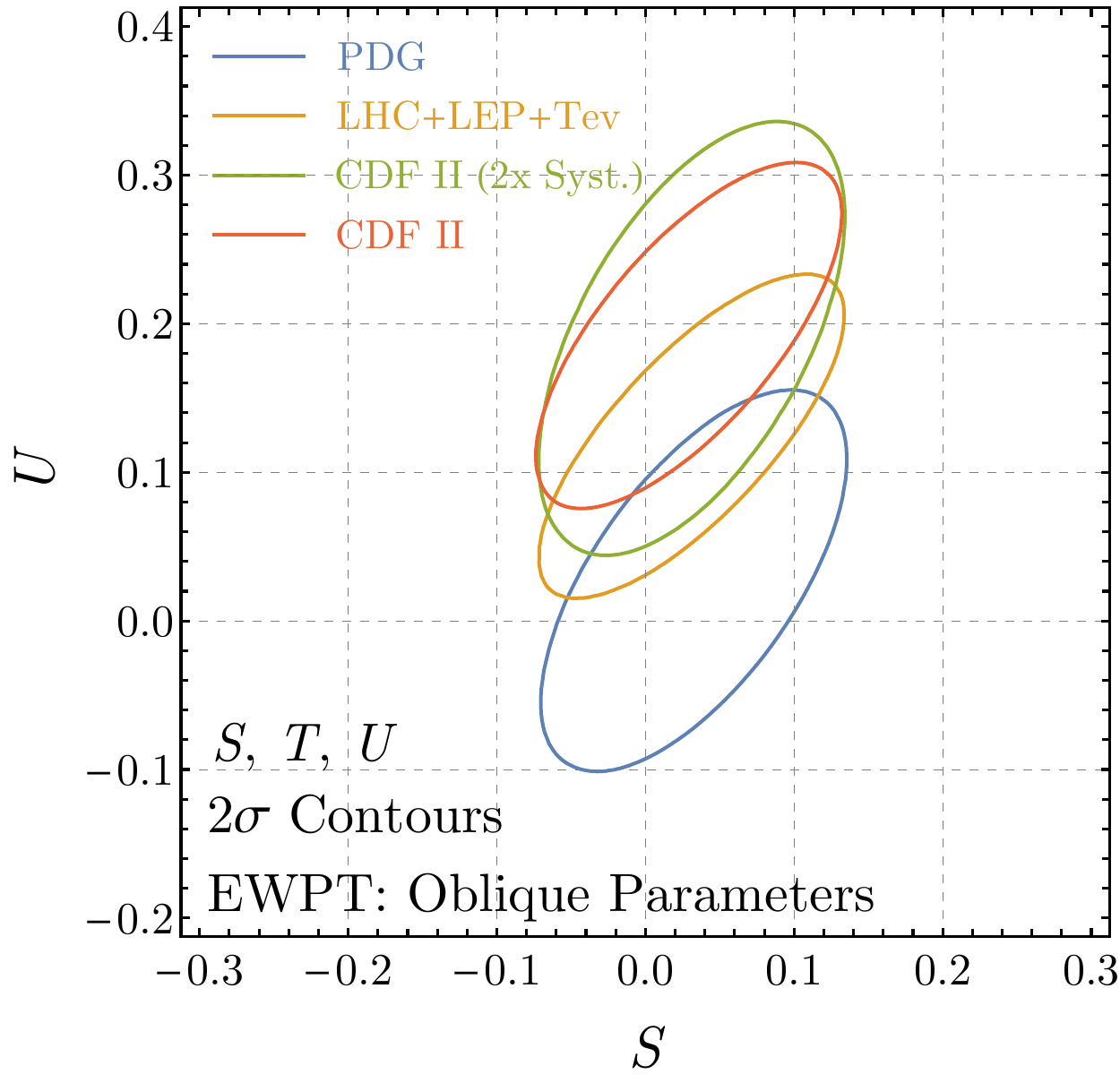}\quad
    \includegraphics[width=.3\textwidth]{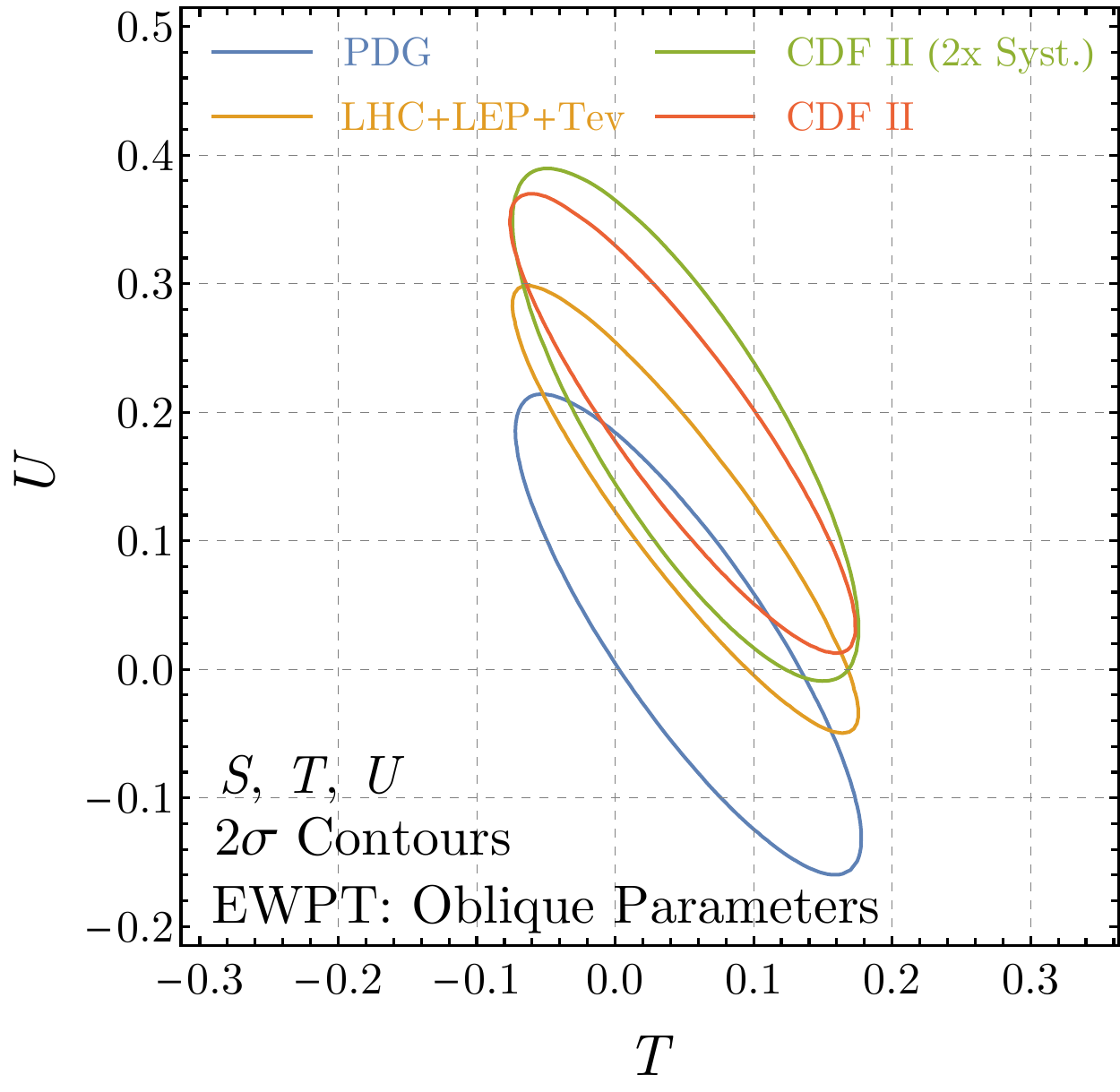} 
    \caption{Similar to \Fig{fig:STplane}, but now also marginalizing over $U$ in the global fit. We show the 2$\sigma$ preferred region of all oblique parameters in the $S-T$ plane (left), $S-U$ plane (center), and $U-T$ plane (right). We find that when we include $U$ in the fit, $S$ and $T$ remain nearly centered about 0, whereas $U$ has a notable positive shift. Getting such large values of $U$ are quite challenging in perturbative models. }
    \label{fig:STplane_Umarg}
\end{figure}

To illustrate this in more detail, we perform the fits to the $S$ and $T$ parameters as described above again, except that we now marginalize over the value of $U$ in addition to the free observables, rather than fixing it to zero. The results are shown in Fig.~\ref{fig:STplane_Umarg}.
We see that, when marginalizing over $U$, the $2\sigma$-preferred range of $S$ and $T$ with the new CDF measurement of $M_W$ is quite similar to the allowed region using the smaller value of $M_W$. Instead, the $U$ parameter is inflated to account for the shift in mass. 

%%%%%

%%%%%

The difficulty in this interpretation is that a large value of $U$ is challenging to generate in perturbative models, because, as mentioned in \Sec{sec:EWPT}, $U$ corresponds to a dimension-8 operator~\cite{Grinstein:1991cd}, and a value of $\mathcal{O}(0.1)$ indicates scales of order $\textrm{few } 100\,\textrm{GeV}$ for tree-level models, and $\ll 100\,\textrm{GeV}$ for particles contributing in loops. 
As the $U$ parameter violates custodial symmetry, it is difficult to imagine a model that generates a large, nonzero value of $U$ without also generating large values of $T$.
We therefore do not attempt to construct models generating large values of $U$. In the concrete BSM models we consider in the next section, we will ignore the (subleading) $U$-dependence altogether.

%============================================================
\section{Implications for BSM Models}
\label{sec:bsm}

From the results of our electroweak fit shown in \Sec{sec:results}, we see that the value of \mw~can dramatically change the preferred values of the oblique parameters. While the 95\% CL region fitting with PDG measurements is nearly centered around the predicted SM values of $(S,T,U) = (0,0,0)$, the updated value of \mw~shifts this region to positive $\mathcal{O}(0.1)$ values of oblique parameters (see \Figs{fig:STplane}{fig:STplane_Umarg}). 

In this section we explore various tree-level and loop-level contributions to the oblique parameters from simple models, and assess their viability. For clarity, we focus on the scenario of \mw~equal to the world average from Tevatron, LEP, and LHC measurements (second scenario in Eq.~\eqref{eq:Wscenarios}). 

It is first worthwhile to estimate the scale of new physics implied by $\mathcal{O}(0.1)$ values of $S$ and $T$. Comparing to the dimension-6 operators defined in \Eq{eq:effOper}, we see that for tree-level matching with perturbative couplings, these operators can be generated by new physics at a scale $\Lambda \sim \textrm{TeV}$.
If the new physics arises in loops, on the other hand, the loop factor suppression implies a scale closer to $\mathcal{O}(100\,\textrm{GeV})$. We will examine this matching in both scenarios, first considering minimal extensions to the SM that can be integrated out at tree-level, such as an additional gauge boson or scalar, then consider a one-loop example with new singlet-doublet fermion pairs.
Note that, as indicated in Fig.~\ref{fig:STplane}, it is important for these models to shift $T$ to positive values to be consistent with our electroweak fit.

\subsection{Tree-Level Models} 

Here we consider models that lead to corrections to the oblique parameters at tree level. Given the results of the fits shown in Fig.~\ref{fig:STplane}, we are particularly interested in models that can accommodate large positive values of $S$ and $T$. 

The simplest examples of models leading to oblique parameter corrections are new vectors that acquire couplings to the Higgs.
As an example, consider a $Z^{\prime}$ from a spontaneously broken additional U(1) gauge symmetry. The oblique corrections from a new gauge boson have been worked out in the most general case in \Ref{Holdom:1990xp}. Focusing on the simplest example, where the gauge boson has only kinetic mixing with SM hypercharge gauge boson, $\mathcal{L} \supset - \frac{1}{2}\epsilon B^{\mu\nu} Z^{\prime}_{\mu\nu}$, the corrections to the oblique parameters are given by
%boe%
\begin{equation}
    S = \frac{4 s_W^2 c_W^2}{\alpha} \epsilon^2 \frac{1}{1 - r} \bigg( 1 - s_W^2\frac{1}{1 - r}\bigg),
    \qquad
    T = - \frac{s_W^2}{\alpha} \epsilon^2 \frac{r}{(1- r)^2}
\end{equation}
%eoe%
where $ r \equiv (m_{Z'}/M_Z)^2$, with $m_Z'$ the mass of the new gauge boson.
While $S$ can change sign depending on whether the new gauge boson is heavier or lighter than $M_Z$, the $T$ parameter in this model is always negative, and therefore cannot resolve the tension in the electroweak fit.

In the more general case with mass mixing, the simple relations above no longer hold, and different values of $S$ and $T$ may be possible, but generating the necessary mixing terms would require strong dynamics involved in electroweak symmetry breaking, which are likely tightly constrained and beyond the scope of this work.

Instead, we are led to consider new scalars affecting the oblique parameters.
An SU(2)$_L$ singlet scalar leads only to an overall rescaling of the Higgs couplings that do not affect $S$ and $T$ or shifts in the Higgs self-coupling. 
Models with extra SU(2)$_L$ doublet scalars, such as a 2HDM~\cite{Branco:2011iw}, can affect the Higgs couplings to the gauge bosons, but these deviations are proportional to $\cos^2(\beta - \alpha)$, the square of the alignment parameter, which from an effective field theory perspective is dimension 8, and therefore cannot affect the oblique parameters $S$ and $T$, which are dimension 6.

An SU(2)$_L$ triplet scalar $\varphi^a$, however, leads to more interesting possibilities \cite{Blank:1997qa}.\footnote{We thank Matthew Strassler for bringing this model to our attention.} 
Such a triplet can have interactions with the SM Higgs $\sim \varphi^a H^{\dagger} \sigma^a H$. 
After electroweak symmetry breaking, this interaction leads to a small vacuum expectation value for the scalar triplet, which shifts the mass of the $W$ bosons without changing the mass of the $Z$, therefore offering a possibility of resolving the tension between the CDF measurement of $M_W$ and the SM expectation. 

For concreteness, we will consider a real scalar SU(2)$_L$ triplet $\varphi^a$ with $Y = 0$ which we will refer to as a swino. The Lagrangian takes the form
\begin{equation}
    \mathcal{L} \supset \frac{1}{2}D_{\mu} \varphi^a D^{\mu} \varphi^a - \frac{1}{2} M_T^2 \varphi^a \varphi^a + \kappa \varphi^a H^{\dagger} \sigma^a H - \eta H^{\dagger}H \varphi^a \varphi^a .
\end{equation}
The constraints on SU(2)$_L$ triplet scalars, including the oblique parameters, have been worked out in \Ref{Khandker:2012zu}, where they include the matching up to one-loop order. At tree-level, the contribution to $S$ from scalar triplets vanishes. The $Y = 0$ swino does, on the other hand, lead to a contribution to the $T$ paramter given by
\begin{equation}
    T = \frac{v^2}{\alpha}\frac{\kappa^2}{M_T^4}
\label{eq:trip_T}
\end{equation}
Importantly, unlike the dark photon model, this contribution is {\em positive} for any value of $\kappa$ and can naturally explain the observed discrepancy in \mw~measurement.\footnote{
\Ref{Strumia:2022qkt} also considered a real triplet extension in light of the new $M_W$ measurement, but it has the wrong sign for $T$.}

One can also consider scalar triplets with $Y = 1$, but these lead to the wrong sign for $T$ at tree level. At one loop, both $Y= 0$ and $Y = 1$ triplets lead to additional corrections to both $S$ and $T$, which can be potentially large and positive, depending on the quartic couplings to the Higgs. We leave a more detailed study of these possibilities to future works.
\begin{figure}
    \centering
    \includegraphics[width=0.5\textwidth]{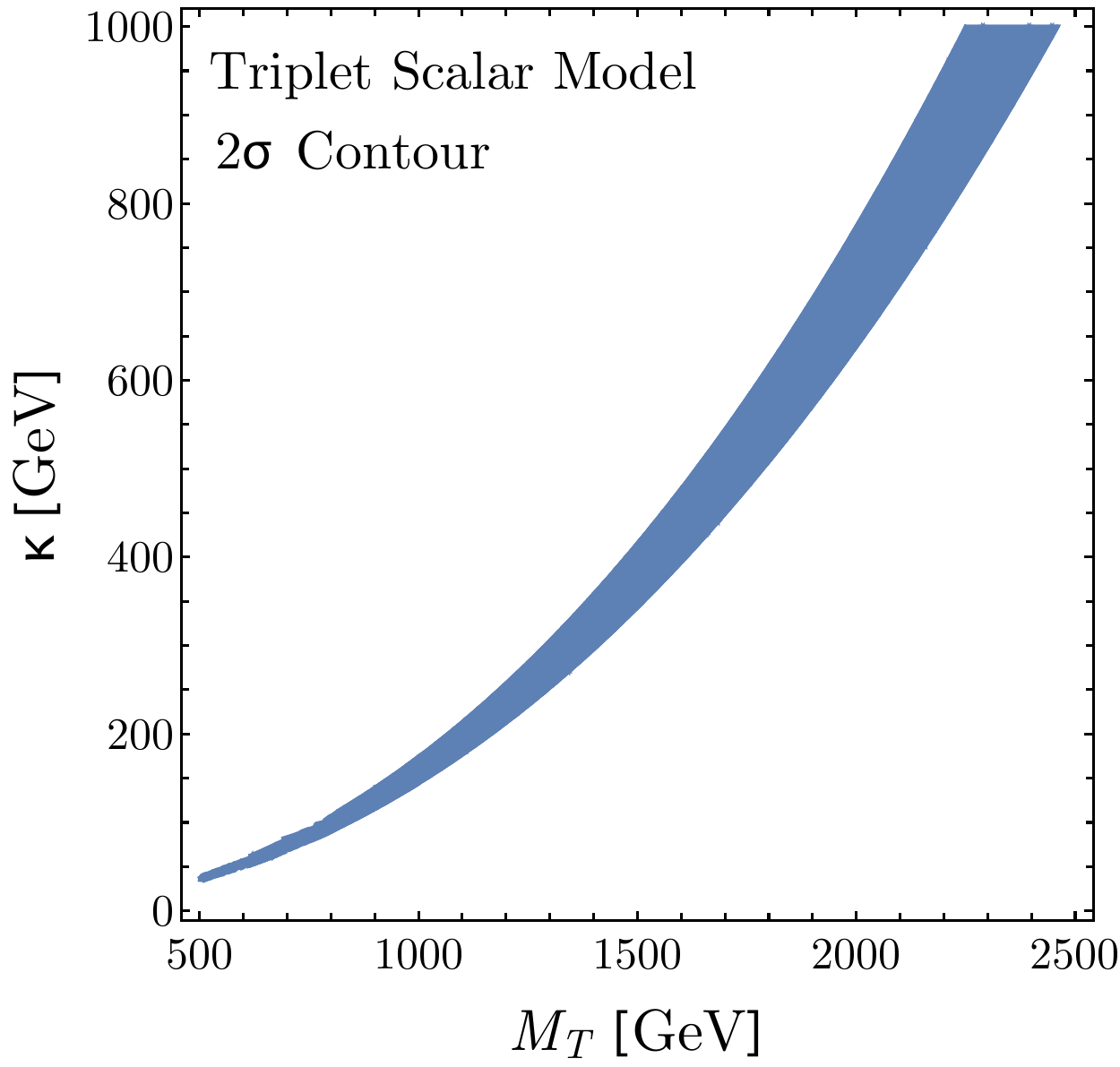}
    \caption{The 2$\sigma$ band in the $M_T-\kappa$ plane of the triplet scalar model. We find viable parameter space for $\mathcal{O}(\mathrm{TeV})$ swino masses that can potentially be probed with future high-energy colliders. }
    \label{fig:tripScalar}
\end{figure}

In Fig.~\ref{fig:tripScalar}, we show the band of values of $\kappa$ and $M_T$ that are compatible with the electroweak fit with the combined value of $M_W$ at $2\sigma$. 
As is clear from the scaling in \Eq{eq:trip_T}, the necessary large value of $T$ can be achieved even for large triplet masses. Requiring $\kappa / M_T \lesssim 1$, the triplet mass can be up to $\mathcal{O}(\textrm{$\sim$ TeV})$, evading any potential collider bounds.

\subsection{Singlet-Doublet Model}
\label{sec:singlet_doublet}

We now shift our attention to another simple extension of the SM, the SU(2)$_L$ singlet-doublet fermion model. Unlike the previous discussion, the contribution of this model to electroweak precision measurements first occurs at loop level. The model includes $N_f$ families of a singlet Majorana and doublet Dirac fermion charged under the electroweak sector~\cite{Mahbubani:2005pt,DEramo:2007anh,Enberg_2007,Cohen:2011ec,Cheung:2013dua,Abe:2014gua,Calibbi:2015nha,Freitas:2015hsa,Banerjee:2016hsk,Cai_2017,Lopez-Honorez:2017ora, Fraser:2020dpy}.\footnote{For simplicity, we consider the scenario where these fermions do not mix with each other, but in principle mixing could lead to richer phenomenology.} 
Such a setup can be embedded inside supersymmetric extensions of the SM. 
The SU(2)$_L$ doublet has hypercharge 1/2 and is composed of two left-handed Weyl fermions $\psi_{2}$ and $\tilde{\psi}_{2}$. The Lagrangian is 
\begin{equation}
\mathcal{L} = \mathcal{L}_{SM} + \mathcal{L}_{\rm kinetic}  -m_2 \psi_{2} \cdot \tilde{\psi}_{2}  - \frac{m_1}{2} \psi_1 \psi_{1} + y \, \mathrm{e}^{i\delta_{\mathrm{CP}}/2} \psi_1 H^\dag \psi_{2} - \tilde y \, \mathrm{e}^{i\delta_{\mathrm{CP}}/2} \psi_{1} H \cdot  \tilde{\psi}_{2}  + \text{h.c.} 
\end{equation}
This Lagrangian has a physical CP-violating phase, as we have four new parameters and three new fields. Since $S$ and $T$ are CP-even observables, we set $\delta_{\mathrm{CP}} = 0$ in this analysis for simplicity. However, this model is also interesting with nonzero values of $\delta_{\mathrm{CP}}$ as it can potentially explain the Galactic Center Excess (see \Ref{Fraser:2020dpy} for details). Additionally, because of the Yukawa terms, there is mass mixing between the fermions and the $\psi_i$ fields are not the propagating degrees of freedom. We call attention to this point because the mass of the lightest propagating fermion is relevant for Higgs (and $Z$) decay constraints, which require $M_\chi > M_h/2$.

\begin{figure}
    \centering
    \includegraphics[width = 120 mm]{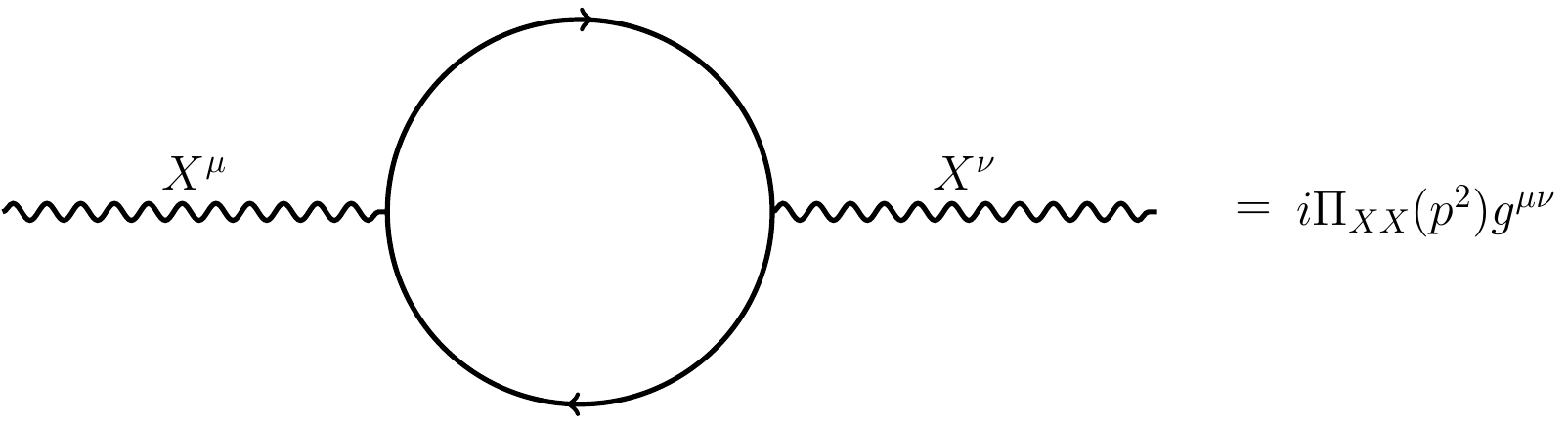}
    \caption{The 1-loop contribution to the vacuum polarization of SM gauge bosons. $X^{\mu}$ here stands for any electroweak gauge boson $X = W, Z, \gamma$.
    }
\label{fig:Oblique_Parameter}
\end{figure}

The singlet-doublet model contributes to the $S$ and $T$ parameters at loop-level with the new fermions running in the loop. While more details of the calculation are given in \Ref{Fraser:2020dpy}, we provide a quick summary here. We write a generic coupling between gauge bosons $i$ and fermions $j$ as $i \gamma^\mu(C_{ij}^V - C_{ij}^A \gamma^5)$ where $C_{ij}^{V}$ and $C_{ij}^{A}$ are the vector and axial vector couplings respectively. In $\overline{\mathrm{MS}}$, we find
\begin{equation}
\begin{split}
i \Pi(p^2) g^{\mu\nu} &= \frac{-i g^{\mu\nu} }{4\pi^2} \int_0^1 dx \left( \left( |C_{Vij}|^2 +|C_{Aij}|^2 \right) p^2 x(1-x) \right.\\
    & \left. + \left( |C_{Vij}|^2 -|C_{Aij}|^2 \right) m_i m_j -  \left( |C_{Vij}|^2 +|C_{Aij}|^2 \right) \Delta \right) \log \frac{\mu^2}{\Delta},
\end{split}
\end{equation}
where $\Delta = m_i^2 + x(m_j^2 -m_i^2)- x(1-x)p^2$. The other relevant expression is $\Pi'(p^{2})$, which is given by
\begin{equation}
\begin{split}
    i\Pi'(p^2)g^{\mu\nu} = \frac{-i g^{\mu\nu} }{4\pi^2} \int_0^1 dx \bigg\{ 2 \left( |C_{Vij}|^2 +|C_{Aij}|^2 \right)  x(1-x)\log \frac{\mu^2}{\Delta} \\
     + \Big[ \big( |C_{Vij}|^2 +|C_{Aij}|^2 \big) p^2 x(1-x) + \left( |C_{Vij}|^2 -|C_{Aij}|^2 \right) m_i m_j   \\  - \left( |C_{Vij}|^2 +|C_{Aij}|^2 \right) \Delta \Big]\frac{x(1-x)}{\Delta}\bigg\}.
\end{split}
\end{equation}

The diagram topology contributing to $S$ and $T$ in this model is shown in Fig.~\ref{fig:Oblique_Parameter} and scales linearly with the number of new fermion generations, $N_f$. 
We can only get a nonzero $T$ value when the custodial symmetry is broken, i.e. $y \neq \tilde y$. Furthermore, $S$ and $T$ both decrease as $m_{2}$ or $m_{1}$ increase, making it difficult to reach values consistent with both the updated electroweak fit and existing experimental constraints. 

\begin{figure}
    \centering
    \includegraphics[width=0.45\textwidth]{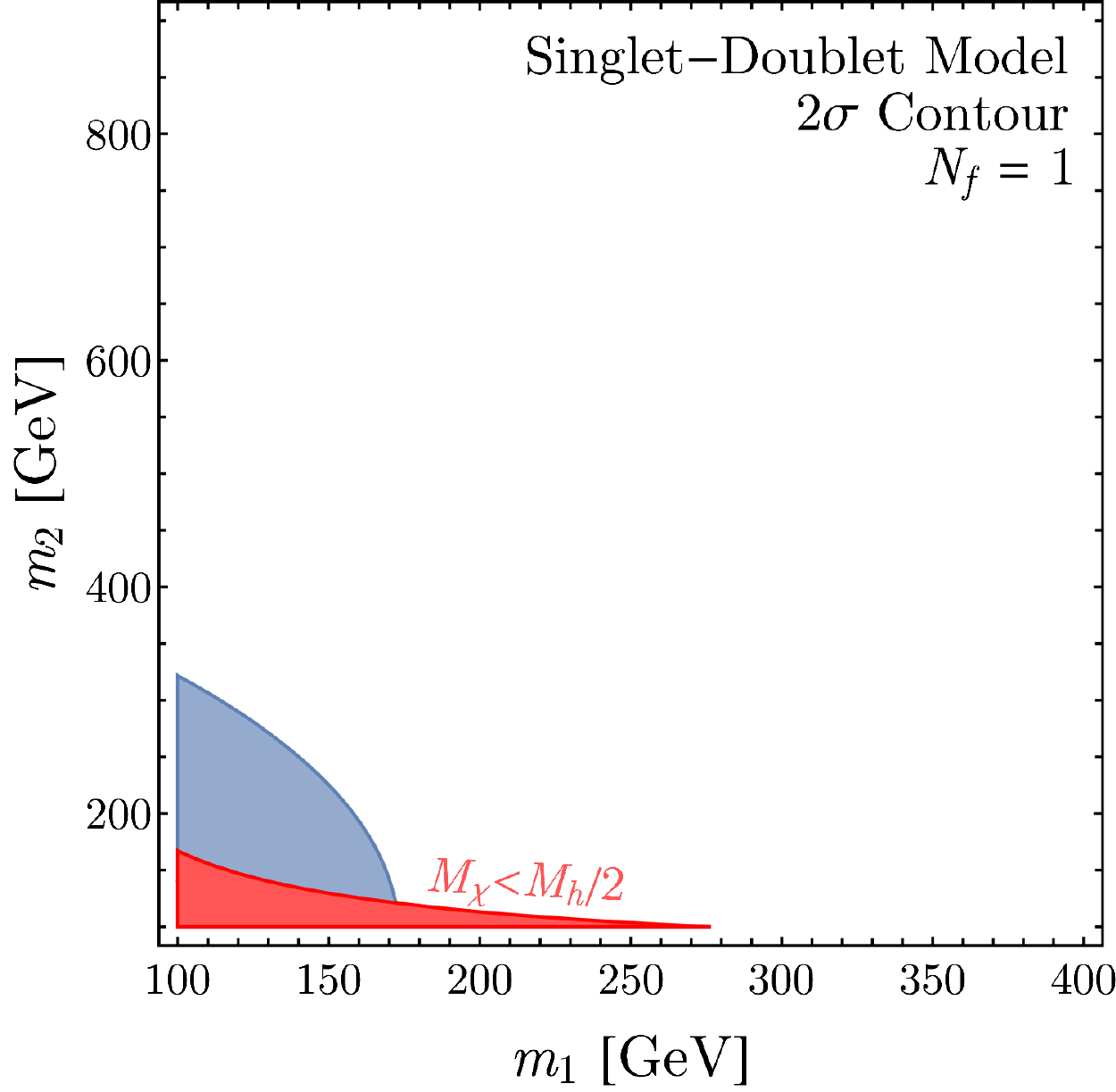} \qquad
    \includegraphics[width=0.45\textwidth]{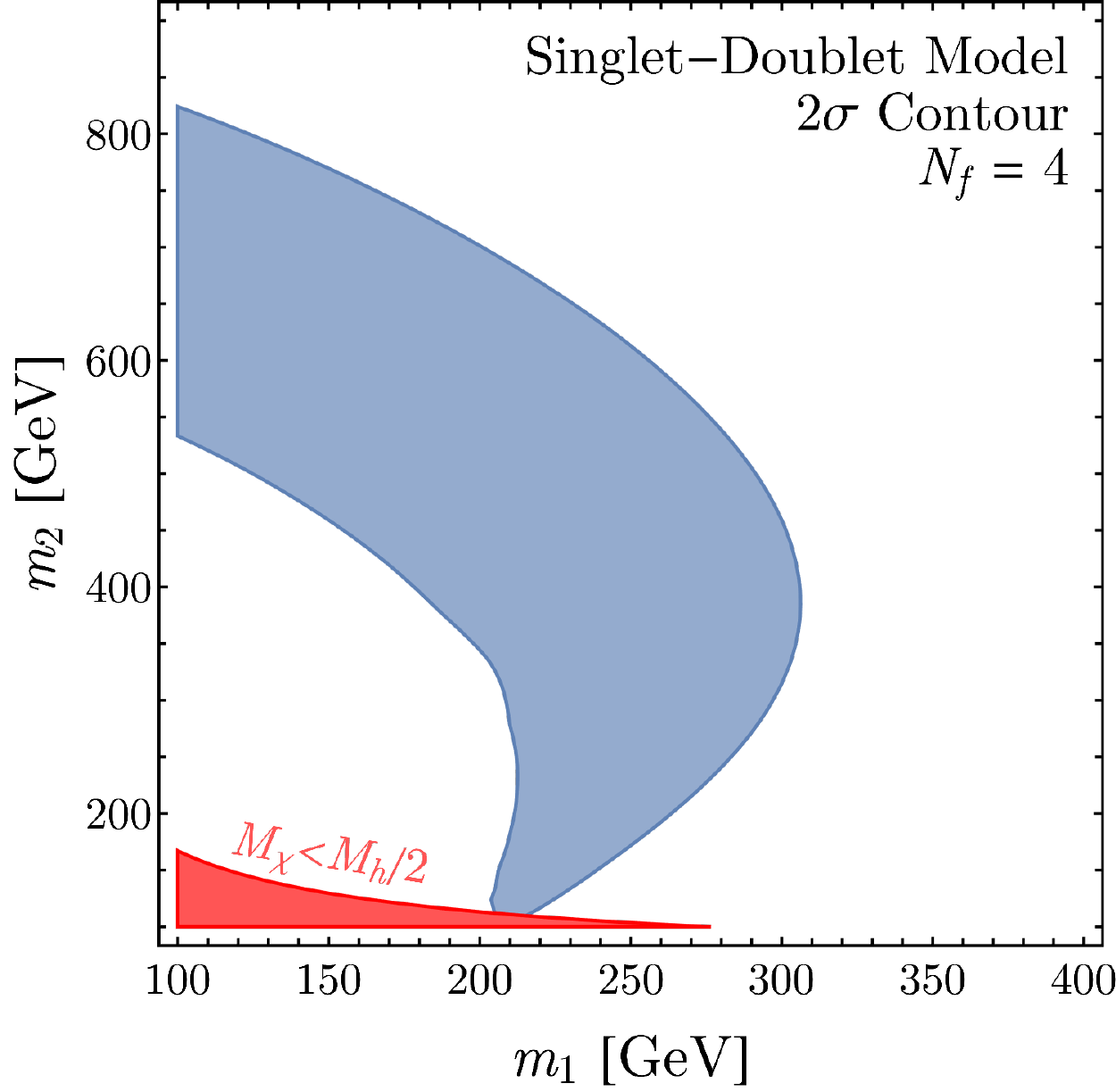}
    \caption{Contribution to the S and T parameters from singlet-doublet fermions for benchmark values of the couplings ($y=0.1,~\tilde{y}=1,~\delta_{\mathrm{CP}}=0$) and $N_f$ generations of new fermions, where $N_f=1$ ($N_f=4$) on the left (right). In the blue band the model can give rise to large enough $S$ and $T$ to explain the world average \mw~measurement within $2\sigma$. Relevant constraints on the model are briefly discussed in the text; in particular, direct LHC searches can potentially rule out most of the blue band for $N_f=1$ and probe much of the $N_f=4$ allowed region.}
    \label{fig:loop_model}
\end{figure}

In Fig.~\ref{fig:loop_model} we plot the dependence of $S$ and $T$ on the new fermion mass parameters $m_1$ and $m_2$ to get a benchmark value of the couplings. 
Lower values of $m_1$ and $m_2$ are strongly constrained by a host of different measurements (including LEP bounds on charged fermions, Higgs and invisible $Z$ decays, and direct searches for light fermions carrying electroweak charge). 
In the left panel of the figure we consider the model with only one generation of new fermions. 
We find that the contribution to $S$ and $T$ is only large enough to explain the CDF~II anomaly in a small corner of the parameter space; direct searches at LHC strongly constrain this range of masses.

In the right panel of Fig.~\ref{fig:loop_model} we show the contribution of the model to the oblique parameters with $N_f=4$. 
We now find a larger range of masses that give rise to \mw~values within $2\sigma$ of the global average measurement. 
Direct LHC searches can again rule out some of this parameter space, but there is still viable parameter space in the range of masses shown in the figure, specifically in the limit of degenerate masses or at high values of $m_2$. 
A more thorough exploration of the viable parameter space (with other values of $y$ and $\tilde{y}$) is left for future work.

\section{Conclusion}
\label{sec:conc}

In this paper we studied the effect of recent \mw~measurement at CDF~II on global fits of electroweak precision observables and the implications for physics beyond the SM. By performing a standard $\chi^2$ fit over SM parameters as well as the oblique parameters $S$, $T$, and $U$, we explored the efficacy of a variety of models for generating an upward shift in the \mw~ mass. After combining all \mw~measurements at the Tevatron, LEP, and the LHC, there exists a significant discrepancy with SM predictions. 

The results of our fit suggest that new physics models that contribute to $S$ and, more substantially, a positive $T$ are potential candidates to explain the anomaly. While we considered a global fit also including $U$, the results did not have a natural model-building interpretation. Of the models we consider, we find that a generic dark photon mixing with SM photon, a singlet scalar extension of SM, and a 2HDM model fail to yield $S$ and $T$ contributions consistent with our fit. However, the swino model was markedly successful since it generated positive $\mathcal{O}(0.1)$ values of $T$ in unconstrained regions of parameter space. Viable triplet mass values were found to be near or above the TeV scale, which can evade current experimental bounds while giving rise to interesting signatures in future high energy colliders such as FCC-hh or muon colliders. We leave a detailed study of such signals to future work. Additionally, we found some success with a singlet-doublet fermion model when considering multiple generations.

As previously mentioned, there are other anomalies in the SM that could arise from discrepant electroweak precision measurements, such as the anomalous magnetic moment of the muon $g-2$. 
It was pointed out in Ref.~\cite{Crivellin:2020zul} that the existing discrepancy between the theoretical and measured values of $(g-2)_{\mu}$ can be absorbed in a shift to the hadronic vacuum polarization contribution by changing $\Delta \alpha_{\textrm{had}}^{(5)}$, at the cost of increasing the tension in the SM electroweak fit, particularly by {\em decreasing} the preferred value of $M_W$. It is of high importance to explore if the necessary change in the fit to ameliorate the $(g-2)_{\mu}$ discrepancy can be accommodated by the BSM effects of interest for the $W$ mass measurement, or if something much more exotic is required.

Finally, we would like to call attention to the fact that a tension arising from the global SM electroweak fit is not unique to the $W$ boson mass. For example, significant deviations from the SM have been evident the forward-backward asymmetry observable at LEP for many years~\cite{ALEPH:2005ab}, and numerous attempts at explaining this with BSM physics (e.g. \Refs{Choudhury:2001hs}, among others). This further motivates future study of how potential new physics affects electroweak precision observables.

These results can be interpreted as new \textit{oblique} signs of BSM appearing around the TeV scale. In light of this new measurement,  further experimental results, including improvement to measurement of \mw~at LHC or future colliders, are strongly motivated.

\vspace{4ex}
\noindent
Note added: As this paper was being finalized, Refs.~\cite{Fan:2022dck, Lu:2022bgw, Athron:2022qpo, Yuan:2022cpw, Yang:2022gvz, deBlas:2022hdk, Zhu:2022tpr} appeared, which also consider the implications of the recent $M_W$ measurement.

\section*{Acknowledgements}

We are especially grateful to JiJi Fan, Matthew Reece and Matthew J. Strassler for their engagement, feedback and suggestions. We are also thankful to JiJi Fan and Matthew Reece for sharing their electroweak fitting code.
We would also like to thank Ayres Freitas, Tao Han, Matthew Low, Juli\'an Mu\~noz, David Shih, and 
Weishuang Linda Xu for helpful discussions. 

The work of PA was supported by the U.S. Department of Energy, Office of Science, Office of High Energy Physics, under grant Contract Number DE-SC0012567. CC and SH are supported by the DOE Grant DESC0013607. CC is also supported by an NSF Graduate Research Fellowship Grant DGE1745303. SH is also supported in part by the Alfred P. Sloan Foundation Grant No. G-2019-12504. Some of the computations in this paper were run on the FASRC Cannon cluster supported by the FAS Division of Science Research Computing Group at Harvard University.

{\small
\renewcommand{\baselinestretch}{1.0}
\bibliographystyle{utphys}
\bibliography{ref}
}
\end{document}